# A Deep Learning Representation of Spatial Interaction Model for Resilient Spatial Planning of Community Business Clusters

Haiyan HAO[1] and Yan WANG[2*]

[1] Ph.D., Department of Urban and Regional Planning and Florida Institute for Built Environment Resilience, College of Design, Construction and Planning, Univ. of Florida, 1480 Inner Rd., Gainesville, FL 32611. Email: hhao@ufl.edu.

[2] Assistant Professor, Department of Urban and Regional Planning and Florida Institute for Built Environment Resilience, Univ. of Florida, P.O. Box 115706, Gainesville, FL 32611 (corresponding author). Email: yanw@ufl.edu.

**Abstract:** The increasing risks posed by adverse events, such as extreme climatic events and pandemics, have negatively impacted the vitality of community businesses. These challenges necessitate resilient strategies to empower community businesses to adapt to the various adversities. For example, scholars have advocated for improved spatial planning to accommodate customers' changing behaviors during adverse events. However, existing Spatial Interaction Models (SIMs) are limited in capturing the complex and context-aware interactions between business clusters and trade areas. To address this limitation, we propose a SIM-GAT model to predict spatio-temporal visitation flows between community business clusters and their trade areas. The model innovatively represents the integrated system of business clusters, trade areas, and transportation infrastructure within an urban region using a connected graph. Then, a graph-based deep learning model, i.e., Graph AttenTion network (GAT), is used to capture the complexity and interdependencies of business clusters. We developed this model with data collected from the Miami metropolitan area in Florida. We then demonstrated its effectiveness in capturing varying attractiveness of business clusters to different residential neighborhoods and across scenarios with an eXplainable AI approach. We contribute a novel method supplementing conventional SIMs to predict and analyze the dynamics of inter-connected community business clusters. The analysis results can inform data-evidenced and place-specific planning strategies helping community business clusters better accommodate their customers across scenarios, and hence improve the resilience of community businesses.

## 1. Introduction

Community businesses play a crucial role in shaping the vitality and viability of communities. Individually, businesses in sectors tied to "third places", such as *retail trade*, *accommodation and food services*, and *arts, entertainment, and recreation*, provide spaces for local residents to gather, socialize, and share information. Scholars have indicated that such "third places" are vital for building social capital and rapport within communities (Jacob, 1961; Oldenburg, 1999; Xiao et al., 2022). In aggregation, the spatial distribution of business clusters can shape patterns of mobility flows, resource allocation, and employment opportunities (Barata-Salgueiro & Cachinho, 2021; Barata-Salgueiro & Erkip, 2014; Jacobs, 1969). Despite the importance, these community business clusters are highly vulnerable to environmental shocks and their precarity has become particularly evident during the pandemic and recent weather and climate events (e.g., Zhang et al., 2009; Chang et al., 2022; Xiao et al., 2022; Fairlie, 2020).

Though associated with high vulnerability, a large number of community businesses are located in hazard-prone locations, such as the storm surge planning zone in coastal cities, which could alter the composition of community businesses during extreme events (e.g., tropical storms and tidal floods). However, many of these coastal cities are still encouraging developments in hazard-prone areas due to the significant commercial benefits (e.g., Lorincz, 2021), which present significant challenges to the planning practices of community business clusters. In response, scholars and practitioners have promoted different resilience strategies to help community businesses prepare for, respond to, and adapt to adversities (Barata-Salgueiro & Cachinho, 2021). For example, the National Institute of Standards and Technology (NIST) has released a comprehensive guide for assessing and enhancing communities, including community businesses, with a focus on upgrading built environments such as reinforcing buildings and infrastructure (Cauffman, 2015). Some scholars have suggested that community businesses develop continuity plans and actively engage in disaster preparedness and response activities (Adekola et al., 2020; Jones & Ingirige, 2008). Furthermore, recent studies have emphasized the role of spatial planning in enhancing the resilience of community business clusters (Kärrholm et al., 2014; Barata-Salgueiro & Cachinho, 2021). In particular, scholars have advocated people-centered planning for shopping areas with a data-driven understanding of customers' dynamic behaviors (e.g., Rao et al., 2022; Rogers & Eckenrode, 2021). Corresponding improvements, e.g., pedestrian-oriented designs, can then be implemented to improve the attractiveness of shopping areas by better tailoring to customers' shopping behaviors.

The interaction between customers and business clusters is conventionally modeled with the Spatial Interaction Model (SIM), which calculates the visitation flows between business clusters and trade areas based on business (cluster) attractiveness and travel costs (Zipf, 1946; Sen & Smith, 2012). SIM and its variations, such as Huff's model (Huff, 1964), multiplicative competitive interaction (MCI) model (Nakanishi & Cooper, 1974), and radiation model (Simini et al., 2012), have been widely used in various urban applications, including travel behavior modeling, store location selection, flow generation, and catchment area estimations (Huff & McCALLUM, 2008; Du & Wang, 2011; Simini et al., 2021; Gu et al., 2024). However, these models are limited in accounting for the complex interactions among business clusters and trade areas that distinct customer groups may perceive business (cluster) attractiveness differently (Reutterer & Teller, 2009). Also, these models have rarely considered contextual situations (e.g., holidays and extreme events) that can influence visitors' perception of business attractiveness over time (Sirohi et al., 1998; Liang et al., 2020).

A few recent studies sshave explored the use of spatio-temporal deep learning models to approximate the conventional SIMs (e.g., Simini et al., 2021; Yeghikyan et al., 2020; Shen et al., 2021). Deep learning models are especially suitable for modeling complex systems as the deep-layered structure and various activation functions allow the model to automatedly learn the implicit, interactive, non-linear, and dynamic dependencies among different variables. However, these deep learning models focus on coarse spatial zones and general transportation flows without differentiating the characteristics of origin and destination zones. They also do not account for the population- and context-variant attractiveness of community business clusters. To address these research gaps, we propose a deep learning model, i.e., the Spatial Interaction Model-Graph AttenTion network (SIM-GAT), to model the interactions between community business clusters and trade areas. The model refers to the conventional SIM, i.e., gravity model, but utilizes the GAT model to enable the inclusion of more complex contextual variables, different transportation modes, and dynamic attractiveness of community business clusters. We showed the effectiveness of the model by calibrating it with data collected from the Miami metropolitan area in Florida, the U.S. We further interpret the calibrated model with eXplainable AI (XAI) and identify key variables influencing the attractiveness of business clusters across different outlet groups and scenarios. This research contributes a method to model the dynamics of community business clusters for resilience planning, which is important given the increasingly complex and challenging urban environments.

## 2. Literature Review

### 2.1. Conventional Spatial Interaction Model and Variables Influencing Business Attractiveness

SIM refers to a broad set of modeling approaches focusing on predicting flows between origin and destination (Sen & Smith, 2012). The gravity model, in analogy with Newton's gravitation law, is the most widely used type of SIM that calculates the population flow between two geographic zones as Equation (1) shows (Zipf, 1946; Sen & Smith, 2012):

$$T_{ij} = \frac{k M_i^{\alpha} M_j^{\beta}}{c_{ij}^{\gamma}} \tag{1}$$

where

$T_{ij}$: the population flow between zone $i$ and zone $j$;

$M_i$ and $M_j$: the mass terms for zone $i$ and zone $j$;

$c_{ij}$: the travel cost between zone $i$ and zone $j$;

$k, \alpha, \beta$: adjustable parameters for calibration; and

$\gamma$: distance decay exponent.

When applied to the modeling of dynamics of community business clusters, the mass terms $M_i$ and $M_j$ in Equation (1) can refer to the demand of the trade area (e.g., measured with population size) and the attractiveness of the business cluster $j$ (e.g., measured with business cluster areas). Extended from SIM, Huff's model (1964) concerns the competitive behaviors among different business clusters and calculates the probability for customer $i$ visiting the business cluster $j$ as:

$$P_{ij} = \frac{A_j{}^\alpha D_{ij}{}^{-\beta}}{\sum_{j=1}^{n} A_j{}^\alpha D_{ij}{}^{-\beta}} \quad (2)$$

where

$A_j$: the attractiveness of business cluster $j$;

$D_{ij}$ the travel cost for customer $i$ to travel to the business cluster $j$; and the

$\alpha, \beta$: adjustable parameters for calibration. $\beta$ refers to the distance decay exponent.

$n$: the total number of business clusters accessible for customer $i$

Previous studies have identified various factors influencing the attractiveness of businesses at individual and agglomeration levels. At the <u>individual</u> level, studies found that *size*, *store image*, *product quality*, *product assortment*, and *(discounting) prices* influence customers' patronage decisions and loyalty (Reutterer & Teller, 2009; Sirohi et al., 1998; Piovani et al., 2017). It was also found that these attractiveness factors can exert different influences across customer groups. For example, wealthy customers are more sensitive to *product quality* while less to *discounting prices* (Talukdar, 2008; Reutterer & Teller, 2009). Note that the varied attractiveness variables described here refer to customers' perceived attributes (Sirohi et al., 1998). In this sense, the *advertisement* and *visibility* can also increase the attractiveness of individual businesses (Anwar & Climis, 2017).

Businesses tend to <u>agglomerate</u> in cities. The agglomerations can be either naturally evolved (e.g., the town center), planned development (e.g., a shopping mall), or established through a hybrid method (e.g., a commercial plaza) (Teller & Elms, 2010). The agglomeration of businesses can reduce customers' searching costs when comparing products and services from different stores or transportation costs by enabling multi-purpose shopping trips (Popkowski Leszczyc et al., 2004; Sevtsuk, 2014). In addition, business agglomerations with diverse tenants also serve as "habitats" for consumers to relax, socialize, and entertain in addition to shopping (Bloch et al., 1994). These "consumer habitats" are supported by easy access (*accessibility*), *shared transportation facilities* (e.g., parking garages), *hospital atmosphere*s, and various food, service, and entertainment *facilities*, all of which can contribute to the attractiveness of business clusters (Sevtsuk, 2014; Teller et al., 2016; Teller & Elms, 2010; Teller & Reutterer, 2008; Hart et al., 2007; Dolega et al., 2016).

Despite the diverse factors influencing the attractiveness of business agglomerations, conventional SIMs only use a single measure, i.e., *size* in most cases, to quantify the attractiveness of business centers. To integrate the diverse attractiveness variables, studies explored multiplication-, addition-, and regression-based approaches (Dolega et al., 2016; Nakanishi & Cooper, 1974; Sevtsuk, 2014; Rinner et al., 2019). For example, the multiplicative competitive interaction (MCI) model extends Huff's model by integrating different attractiveness variables with cumulative multiplication (Nakanishi & Cooper, 1974). Addition-based methods combine attractiveness-relevant variables of business clusters indiscriminately (Dolega et al., 2016). Regression-based models assign weights to different attractiveness variables by fitting the empirical data (Sevtsuk, 2014). Moreover, studies have shown that the attractiveness of individual and agglomerated businesses are shaped by customers' perceptions that may vary among customer groups and across situations (e.g., holidays or extreme events) (Reutterer & Teller, 2009; Sirohi et al., 1998; Hart et al., 2007). These additions-, multiplication-, and regression-based methods for measuring business attractiveness yield static estimations that do not change among customer groups and situations.

With respect to the travel costs in SIMs, existing studies tended to measure them in terms of time or distance (Yiannakoulias et al., 2013). For spatial distances, Yue et al. (2012) showed that the network distance can yield a more realistic result compared to the Euclidean distance in the calibration of SIM. Al-Hashimi & Mansour (2014) have compared three types of spatial distance, including metric, angular, and topological distances, for calibrating Huff's model. The results showed that the model calibrated with topological distance yielded the best performance. The temporal distance, which can vary greatly within a day according to traffic conditions, is often used in traffic assignment models for transportation planning (Hellman, 2010). Some researchers also proposed approaches to calculate travel costs for specific travel modes, e.g., walking and transit (Sevtsuk, 2021; Tahmasbi et al., 2019). Levinson & Kumar (2008) integrated the multi-modal travel choices (e.g., walking, transit, and automobile) in the formulation of travel costs with a weighted summation. Other studies (e.g., Geurs et al., 2016) also formed generalized travel costs with regression models to account for the varying travel modes and multi-variate influential factors, which give more flexibility and generalizability to the SIM. Despite the various approaches, most studies still tend to form an individual measure for travel costs. Recent research has seen substantial advances in the application of graph theory in representing multimodal urban transportation networks, but its potential for modeling the dynamics of community business clusters has not been explored yet.

### 2.2. Empirical Advancements in Spatial Interaction Model

Following the proposal of the various SIMs, numerous empirical studies have been conducted to validate and calibrate the models with travel surveys or big data sources. Notably, big data, such as GPS tracks and social media check-in records, provide more extensive and fine-grained data and hence enable more accurate calibration of SIMs compared to traditional travel surveys. For example, Yue et al. (2012) calibrated Huff's model with vehicle GPS trajectories for seven shopping centers in Wuhan, China. Liu et al. (2014) fit the gravity model with inter-city mobility data mined from social media check-in records and revealed a weaker distance decay effect for inter-city trips compared to intra-city movements. Siła-Nowicka & Fotheringham (2019) presented a method for processing GPS tracks into origin-destination flow matrices, which can be sequentially used for calibrating SIMs. Additionally, apart from the physical interactions captured through human mobility, some researchers have also calibrated SIMs to model human interactions in cyberspace using phone-call and microblog data (Gao et al., 2013; Wang et al., 2018).

These empirical studies have also identified the limitations of conventional SIM approaches and explored solutions. One limitation is that conventional SIMs model pair-wise interactions and ignore spatial dependences and correlations in the data (LeSage & Pace, 2008). Therefore, some researchers have integrated the gravity model or Huff's model with geographically weighted regression, enabling the locally calibrated models (e.g., Suárez-Vega et al., 2015). Other research also integrated the Huff model with space syntax theory to take the graph structure of urban fabrics into consideration (Al-Hashimi & Mansour, 2014). Additionally, conventional SIMs yield static estimations about the modeled parameters and assume that the captured relations retain across time. To lift this limitation, Liang et al. (2020) proposed a time-aware dynamic Huff model enabling time-aware calibration of Huff's model with spatio-temporal big data. Another limitation is that SIM is often calibrated via Ordinary Least Squares (OLS) regression, which assumes that the response variables are normally distributed. However, existing empirical studies measured regional interactions with visitation frequencies, traffic flows, and number of check-ins, which are all count data following Poisson distribution (Zhang et al., 2019). To address this limitation, some studies have proposed to integrate Poisson regression or negative binomial regression to calibrate SIM (e.g., Rui et al., 2016; Zhang et al., 2019).

In addition to modeling the global interactions between regions, some researchers also integrated SIM with agent-based modeling (ABM) to model individual customers' patronage to business clusters. For example, Du & Wang (2011) used ABM to model the relation between businesses and customers. Variables, such as *product diversity*, *discount bonus*, *parking convenience*, *customer loyalty*, *household demographic profile*, and *travel distance/expenses*, are considered in the formulation of customers' patronage preferences. Zhang & Robinson (2022) incorporated Huff's model in ABM to capture the competition among business clusters serving the same catchment areas.

### 2.3. Integrating Spatial Interaction Model with Deep Learning and eXplainable AI

A few recent studies used deep learning approaches to approximate SIMs. For example, Simini et al. (2021) implemented the gravity model with feed-forward neural networks for flow generation. Yeghikyan et al. (2020) integrate SIM and graph convolutional network (GCN) to model human mobility flows. Compared to conventional numeric or statistical models, the use of deep learning models accounts for the implicit and complex relations between various variables, including land use, road network, and various facilities, in representing the attractiveness of destinations. Shen et al. (2021) integrated the gravity model and a convolutional auto-encoder to predict the near-future passenger flows of metro stations in Beijing, China. Particularly, they calibrated the convolutional neural networks (CNNs) to extract and recover the sparse origin-destination matrices storing passenger flows. Compared to conventional numeric or statistical models, the use of deep learning models accounts for the implicit and complex relations between various variables, including land use, road network, and various facilities, in representing the attractiveness of destinations. Variables such as surrounding Point-of-Interests (POIs) are also included in the model. However, these models are not tailored for community business clusters. The analysis units are metro stations or census tracts that do not delineate business clusters, trade areas, and transportation networks, which are essential in the spatial planning of business clusters.

Though associated with great potential, deep learning models are known for their high complexity and low transparency, leaving users unclear about how the models work and, thus, are less confident about the modeling results. One approach to lift this limitation is eXplainable AI (XAI) which explains how AI models arrive at specific outputs (Murdoch et al., 2019; Gunning et al., 2019; Arrieta et al., 2020). XAI has gained increasing popularity in various domains including the field of urban studies. Scholars have suggested that XAI can support smart city applications, such as smart healthcare, finance, transportation, and criminal justice, by examining the validity of AI model outputs (Javed et al., 2023). Furthermore, XAI has been employed to examine the effectiveness of urban management and planning. For example, Chew & Zhang (2022) applied the SHapley Additive exPlanations (SHAP) method to examine the effectiveness of government control measures in containing the COVID-19 pandemic. In Dou et al. (2023), the authors used SHAP to interpret the modeling results for housing prices in Shanghai, China and showed the non-linear influences of various explanatory variables. Hao & Wang (2023) used DeepLift to examine planning factors contributing to community resilience in facing extreme weather events. In Amiri et al. (2021), Local Interpretable Model-agnostic Explanations (LIME) was used to interpret the predictions of households' transportation energy uses. Additionally, Kim et al. (2023) also used SHAP to explain factors contributing to urban expansion. These explanation results can inform urban planning and management practices to support the development of more resilient, safe, and sustainable urban environments.

Inspired by these previous studies, we propose the SIM-GAT model that integrates the conventional SIM with a graph-based deep learning model, i.e., GAT, to model community business cluster dynamics. The model learns the complex and implicit interactions among variables depicting the attractiveness of business clusters and demands of trade areas with real-world customer visitation data. The detailed model architecture and developments are introduced in the following sections.

### 3. Methods

This section introduces the components of the proposed SIM-GAT model. We first reduce the system of community business clusters, consisting of business clusters, trade areas (i.e., residential neighborhoods), and transportation infrastructure, into connected graphs with nodes and edges. We then develop the SIM-GAT model based on the graph representation of the system of community business clusters.

### 3.1. Graph Representation of Community Business Clusters at the System Level

Graph theory was used to model complex systems by representing the system with an abstract format, i.e., nodes and edges (Lu & Yang, 2022). This graph representation can effectively capture the interactions among different system components. In the proposed method, we started with constructing graphs referring to the system of business clusters within the study region. The nodes in the graph represent business clusters and residential neighborhoods while the edges represent the transportation connections between the residential neighborhoods and business clusters (**Table 1**). The nodes representing residential neighborhoods are characterized by a set of variables that may influence residents' patronage preferences (e.g., socio-economic conditions and car ownership). The nodes representing business clusters are characterized by variables that influence the attractiveness of business clusters (e.g., size, tenant diversity, and surrounding transportation facilities). The selection of these variables consults before studies reviewed in Section 2.1. For the multi-modal transportation connections, we represent them with numeric matrices storing the travel costs between each pair of business cluster nodes and residential neighborhood nodes. Additionally, we also represented the dynamic environmental conditions (e.g., weather and holidays) that can impact the performance of community business clusters with a numeric vector. The detailed definitions of nodes, edges, and their assemblies are shown in **Table 1**.

**Table 1**. The graph representation of the system of community business clusters

| Components | Graph representation | Numeric representation |
|---|---|---|
| A business cluster |  Node | A numeric vector $u_i \in \mathbb{R}^l$ consisting of $l$ numeric features describing the "attractiveness" of the business cluster $i$. |
| A residential neighborhood |  A node | A numeric vector $v_j \in \mathbb{R}^k$ consisting of $k$ numeric features describing the demands and patronage preferences of the residential neighborhood $j$. |
| A transportation connection | An edge | A numeric variable $c_{i,j}$ representing the travel cost between residential neighborhood $i$ and business cluster $j$. The transportation network for the whole business cluster system is then represented with matrix $c \in \mathbb{R}^{m*n}$, where $m$ is the number of residential neighborhoods and $n$ is the number of business clusters. For multi-modal systems, multiple matrices can be constructed to store the travel costs for different travel modes (e.g., transit, walk, and automobile). |
| Environmental condition | Global variable | A numeric vector $w_t \in \mathbb{R}^s$ suggesting the environmental conditions (e.g., weather, extreme event, and holidays) of the system of community business clusters at time $t$. |
| A business cluster and its catchment areas | 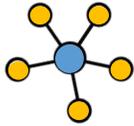 Graph clique | A subgraph $G(u_i, v, e_i)$, where $u_i$ is a commercial center, $v \in \mathbb{R}^{m*l}$ is the set of residential neighborhoods within the catchment area of the business cluster $u_i$, and $e_i$ are the set of edges between residential neighborhoods $v$ and the business cluster $i$. |
| An integrated system of community business clusters | 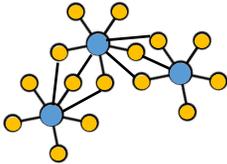 Graph | A graph $G(u, v, e)$, where $u \in \mathbb{R}^{n*l}$ and $v \in \mathbb{R}^{m*k}$ are sets of business clusters and residential neighborhoods. $e$ are the set of edges that connect the two types of nodes. |

### 3.2. Architecture of the SIM-GAT model

We adapt the conventional SIM into a deep learning format with the Graph AttenTion network (GAT) (Veličković et al., 2017). We choose GAT over other deep learning models for the following reasons. First, as we have abstracted the system of business clusters within an urban area as a connected graph (**Table 1**), GAT is a graph-based deep learning model for analyzing graph-structured data. Second, different from other graph-based deep learning models, GAT does not pre-assume the graph structure, i.e., the number of nodes and their connectedness. Instead, GAT calculates the weight of each edge, e.g., $e_{i,j}$, with the attributes of nodes $n_i$ and $n_j$ during the learning process. For this study, readers can view the graph representation of the integrated system of business clusters as a fully-connected graph at the start, whose edge weights are calculated with the attributes of business clusters, residential neighborhoods, and environmental conditions. This calculation, however, will yield abundant edges with weights to be zero or close to zero (e.g., due to low business cluster attractiveness or large travel costs), making the graph look sparse. This setting allows the "attractiveness" of a business cluster to vary across residential neighborhoods with distinct characteristics, as well as varying with environmental conditions by including the environmental vector, i.e., $w_t$ defined in **Table 1**, in the calculation of attention scores (Equation 3). Consequently, the trade area of a business cluster is also dynamic and can be determined in a posterior manner with a threshold, e.g., neighborhoods contributing to 80% of customer flows. Additionally, as GAT learns the localized interactions and does not pre-assume the graph structure, it can be adapted to unseen graphs with arbitrary numbers of nodes after model development (Veličković et al., 2017).

Given a city with $m$ residential neighborhoods served by $n$ business clusters. Each business cluster can take customers from the $m$ residential neighborhoods. In contrast to the multiplication used by the gravity model (Equation 1), SIM-GAT uses the attention mechanism to capture the interactions between each pair of business clusters and residential neighborhoods conditioned by environmental exposure (Equation 3). Specifically, the inputs, including matrices storing features describing business clusters (i.e., $u \epsilon \mathbb{R}^{n*l}$) and residential neighborhoods (i.e., $v \epsilon \mathbb{R}^{m*k}$) are firstly projected to the same hidden dimension via embedding layers (**Figure 2**). The sequence of environmental conditions (i.e., $w_{t-T+1:t} \epsilon \mathbb{R}^{T*s}$) is processed with an LSTM layer and is also projected to the same hidden dimension (**Figure 2**). The resulting matrices and vectors are then broadcasted to the same dimension (i.e., $\mathbb{R}^{m*n*h}$) and concatenated as the input ($\mathbb{R}^{m*n*3h}$) to the graph attention layer (**Figure 2**). The *attention* layer trains a learnable weight vector (i.e., $\vec{a} \epsilon \mathbb{R}^{3h}$ in Equation 3) to produce attention scores for each node pair by taking the dot production with the concatenated representation of node pairs and environmental conditions (Equation 3) (**Figure 2**). Therefore, for each residential neighborhood (e.g., $v_j$), the *attention* layer yields $m$ attention scores (i.e., $e_{1j}, e_{2j}, \ldots\ldots e_{mj}$) suggesting the attractiveness of the $m$ business clusters to that neighborhood. The learnable vector $\vec{a}$ is updated through the training process to assign higher attention scores to business clusters that are more attractive to residential neighborhood $j$. Inspired by the gravity model, we modified the resulting attention score $e_{i,j}$ by the travel cost (i.e., $c_{i,j}$) between the node pairs (Equation 4). In this study, the travel cost matrices $c \epsilon \mathbb{R}^{m*n}$ are pre-computed based on the transportation infrastructure of the studied area.

The modified attention score $\tilde{e}_{i,j}$ is normalized with the SoftMax function as Equation (5). The calculated attention scores operate as distributing the demand for each residential neighborhood to different business clusters (Equation 5) (**Figure 1a**). Then, the hidden state of the business cluster $i$, i.e., $h_i$, can be computed by aggregating the information from the different residential neighborhoods as Equation (6) and **Figure 1b** shows. This step works as aggregating the information of the distributed demands from different residential neighborhoods to the business cluster.

$$e_{i,j,t} = activation(\vec{a}^T[u_i || v_j || w_t]) \tag{3}$$

$$\tilde{e}_{i,j,t} = \frac{e_{i,j,t}}{c_{i,j}} \tag{4}$$

$$\alpha_{i,j,t} = \frac{\exp(\sigma(\tilde{e}_{i,j,t}))}{\sum_i \exp(\sigma(\tilde{e}_{i,j,t}))} \tag{5}$$

$$h_{i,t} = \sum_{j=0}^{m} \alpha_{i,j,t} W_v v_j \tag{6}$$

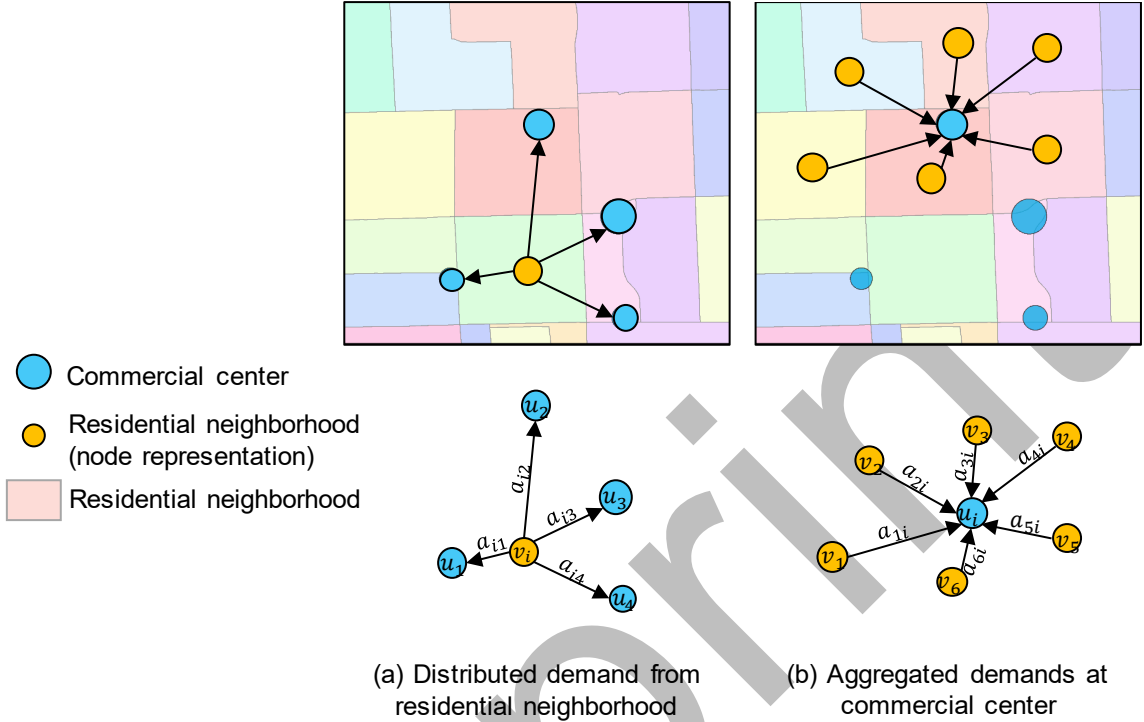

**Figure 1.** The illustration of graph attention used for the system of community business clusters

Equation (3)-(6) can be achieved with a single GAT layer (Veličković et al., 2017). Compared to the original GAT proposed by Veličković et al., (2017), we make a few necessary revisions in the SIM-GAT model: 1) the calculation of attention scores is extended with environmental exposure $w_t$ and then modified by pre-computed travel costs as Equation (3) and Equation (4) shown.; 2). the graph includes two types of nodes, defined in **Table 1**, denoting business clusters and residential neighborhoods neighborhood respectively; The SoftMax is applied to different business cluster nodes resembling their competitions while the aggregation is applied to residential neighborhood nodes. For tasks that output visitation flows, the final step can be achieved by replacing Equation (6) with Equation (7); and 3). an exponential activation is used in the last layer of model output considering that the outputs, i.e., visitations, are count data following Poisson distribution. **Figure 2** shows the overall architecture of the proposed SIM-GAT model. For cases with more than one travel cost matrix (e.g., calculated for different travel modalities), a learnable linear layer is used to combine the different matrices into a single one, which resembles the weighted summation used by previous studies (e.g., Levinson & Kumar, 2008).

$$h_{i,j,t} = \alpha_{i,j} W [u_i || v_j || w_t] \tag{7}$$

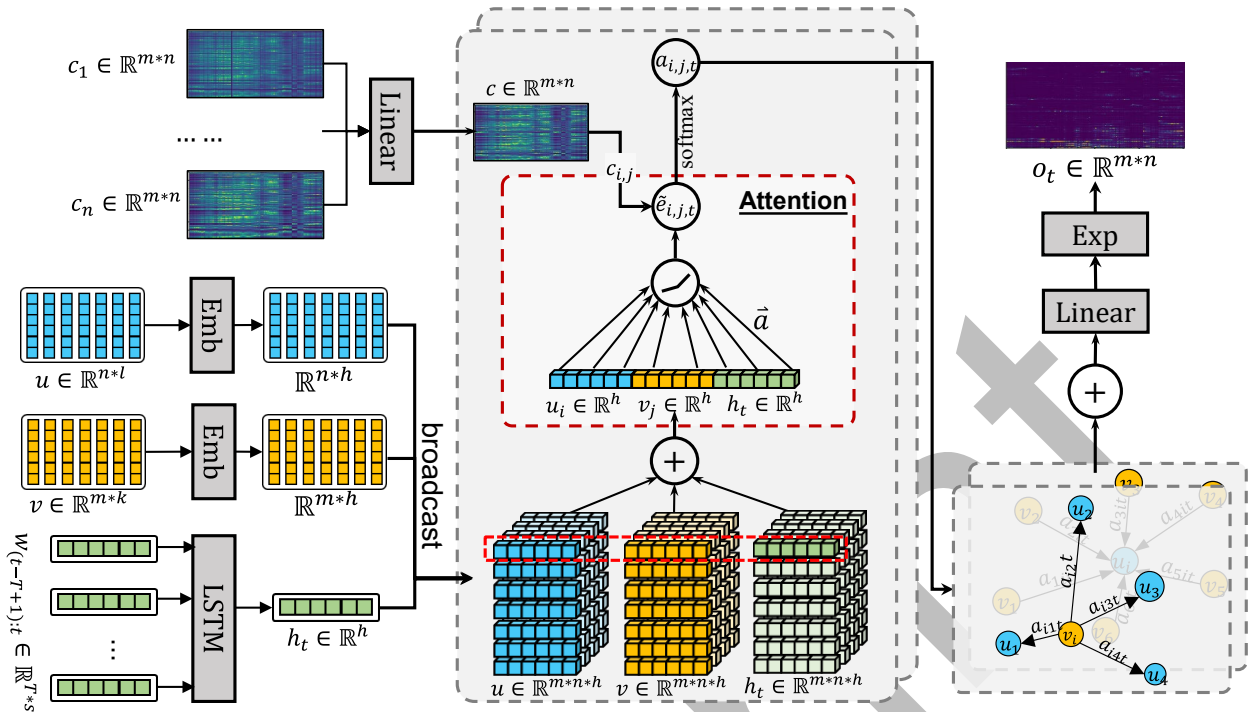

**Figure 2**. The architecture of the SIM-GAT model.

## 4. Case and Data

We chose the Miami metropolitan area in Florida as the study case to calibrate and evaluate a place-specific model. We made this choice primarily considering the following two reasons. First, Miami is located on the U.S. Southeast Coast, which is prone to recurrent environmental shocks such as coastal flooding and tropical storms (Woetzel et al., 2020). A preliminary investigation conducted by us has identified 2,189 businesses in sectors relying on visitations to sustain located in Storm Surge Planning Zone A and B in Miami, where residents are supposed to evacuate during hurricanes. However, continuous commercial and mixed-use developments exist in these hazard-prone areas (Lorincz, 2021). Knowing how business clusters perform across hazard scenarios can inform proactive resilience planning efforts to help community businesses cope with adverse events and, consequently, benefit the social and economic well-being of this region. Second, Miami is among the most socially segregated regions in the U.S. The city includes neighborhoods dominated by socially-vulnerable residents (Othering & Belonging Institute, 2021). It is interesting to investigate how attractiveness of business clusters can vary with neighborhoods' socio-demographic characteristics. We collected data from the study region spanning a two-year period, from January 1, 2019, to November 17, 2020, for model development and validation.

### 4.1. Collect and pre-process the point of interest (POI) and visitation data

We collected the business and visitation data from the *Core Place* dataset and *Pattern* dataset provided by SafeGraph (Safegraph, n.d.a & n.d.b). The *Core Place* data include more than 6 million unique POIs across the United States (Safegraph, n.d.a). Each POI record includes a unique ID, spatial coordinates, North American Industry Classification System (NAICS) code, and other attributes. We used ArcGIS's spatial join tool and identified 24,708 POIs located within the study region. In this study, we used visitation counts provided by the *Pattern* dataset to measure business attractiveness, given that customers can "vote with their foot". This assumption may be less applicable to certain business categories (e.g., manufacturing and construction). Therefore, we focused on business clusters primarily comprised of businesses in *Retail Trade* (NAICS 44 and 45), *Arts, Entertainment, and Recreation* (NAICS 71), *Accommodation and Food Services* (NAICS 72), which compete with each other for customers' foot traffic to thrive and thus are more vulnerable to extreme events or pandemic that limit residents' foot traffic. These three categories consist of around half (i.e., 114,77) of the total POIs in the study region.

The *Pattern* dataset provides the visitation counts directed to the different POIs in addition to visitors' home census tracts, median travel distances, and binned dwell time (Safegraph, n.d.b). SafeGraph collects visitation data through smartphone applications that cooperate with the platform. The panel of the platform includes 47 million devices across the U.S., which accounts for around 10% of the national population and is, therefore, sufficient to uncover urban residents' daily activity patterns.

### 4.2. Identifying and characterizing business clusters

With the collected business data, we used <u>D</u>ensity-<u>B</u>ased <u>S</u>patial <u>C</u>lustering of <u>A</u>pplication with <u>N</u>oise (DBSCAN) to identify commercial clusters. DBSCAN is chosen here because it is robust to noises and can identify clusters of arbitrary shapes, which accounts for the fact that business clusters can take different morphological forms. For example, shopping malls are physically enclosed; commercial plazas are open-air and located in a single or across a few street blocks; while commercial ribbons extend along streets (**Figure 3**). Equation (8) defines the $\varepsilon$-neighborhood of a business $x \in D$, where $dist(x, y)$ is the Euclidean distance between two businesses. DBSCAN identifies cluster seeds for expansion if the number of businesses within $\varepsilon$-neighborhood exceeds the threshold $minPts$.

$$N_\varepsilon(x) = \{y \in D \mid dist(x, y) \leq \varepsilon\} \quad (8)$$

We set $\varepsilon = 200m$ and $minPts = 5$. Therefore, the minimum requirement for a location to be considered as a business cluster is to have more than five businesses in the three NAICS sectors (i.e., *Retail Trade*; *Arts, Entertainment, and Recreation*; and *Accommodation and Food Services*) located within a 200m radius region. These thresholds were selected through a trial-and-error process and based on the trade-off between accuracy and granularity. Specifically, we mapped the clusters, created with different combinations of thresholds (i.e., 100m, 200m, 400m for $\varepsilon$ and 5, 10 for $minPts$) on ArcMap, manually compared the clustering results, and selected the one that can effectively identify business clusters of different sizes and of varying morphological structures. This threshold could be adjusted in future studies for other cities. The DBSCAN initially identified 218 business clusters, which were then manually refined according to built environment features. For example, clusters located on the same street segment were merged as a single cluster. Finally, this process resulted in 199 business clusters in the study region, covering a total of 8,309 businesses in the three sectors. **Figure 4** shows the spatial distribution of business clusters.

After detecting the business clusters, we collected data depicting the business compositions, built environments, and hazard exposure for each business cluster (**Table 2**). These variables are either identified to influence the attractiveness of business clusters by previous research or may differentiate the business cluster's performances across scenarios.

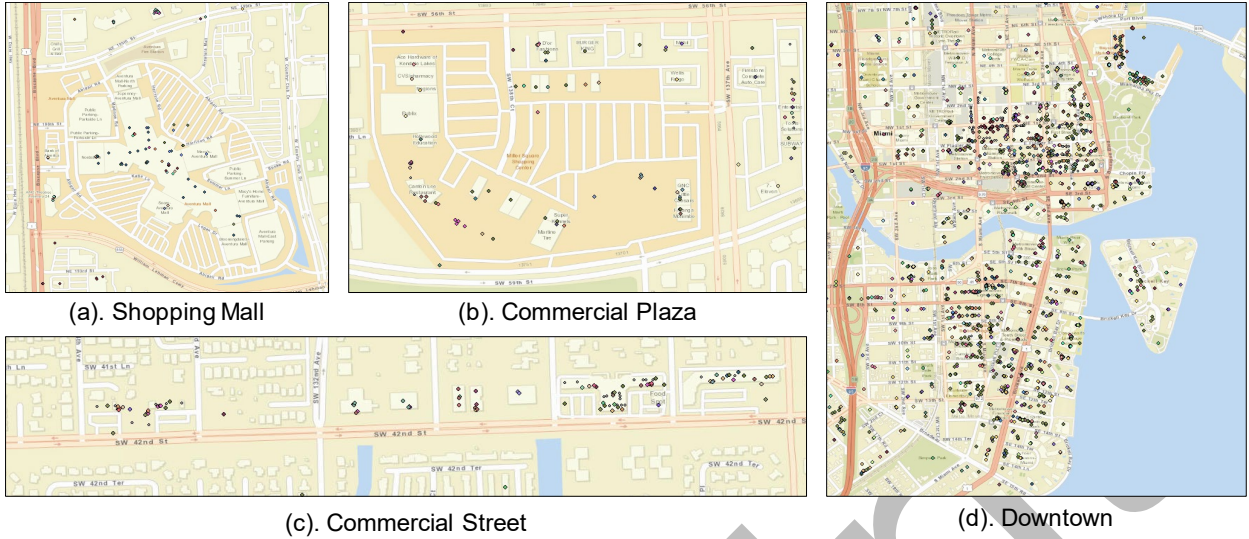

(a). Shopping Mall   (b). Commercial Plaza

(c). Commercial Street   (d). Downtown

**Figure 3**. Examples of business clusters of different morphological structures identified with DBSCAN.

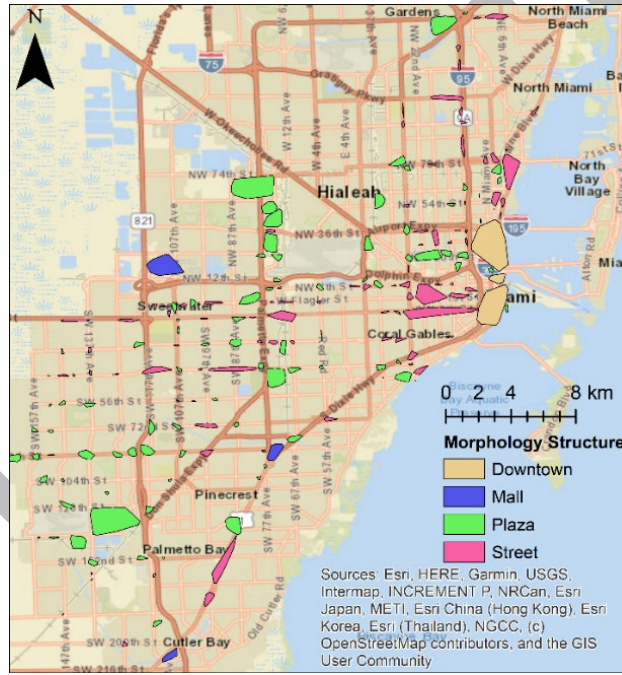

**Figure 4**. The spatial distribution of different business clusters.

**Table 2**. Variables used to characterize business clusters.

| Variable | Definition |
|---|---|
| Morphological structure | A 4-d one-hot vector measuring the morphological structure of the business clusters. The vector takes one of the following values based on manual review:<br>• Commercial plaza;<br>• Commercial street;<br>• Downtown; |
| | • Shopping mall. |
| POI counts | A long vector storing the number of POIs for different NACIS sectors within the business clusters.<br><br>For Miami, this initially results in 126 categories of POIs (at the 6-digit NACIS level). We used Principle Component Analysis (PCA) to project the high-dimension vector into a 7-d vector, which explains the 96.51% variance of the data. |
| POI diversity | A numeric variable measuring the Shannon diversity of POIs, characterized by the 6-digit NACIS code, within the business cluster. |
| The ratio of chain business | A numeric variable measuring the ratio of chain businesses within the business cluster. Chain businesses may have more capacities to counter adverse events and may serve as the anchor stores |
| Land use | An 8-d vector suggesting the proportion of areas of different land uses in the surrounding (i.e., 200m buffer area) of the business cluster including:<br><br>• Commercial;<br>• Mixed;<br>• Office;<br>• Parking;<br>• Recreation;<br>• Single-family residential;<br>• Multi-family residential;<br>• Vacant |
| Transportation facility | A numeric variable measuring the number of bus stops located within or nearby (i.e., 200 m buffer area) the business cluster. |
| Hazard exposure | A 3-d vector suggesting the proportion of areas of different flood zone types around (i.e., 200 m buffer area) the business cluster including:<br><br>• X (low-risk 500-year floodplain);<br>• A (high-risk 100-year floodplain);<br>• V (coastal 100-year floodplain). |

### 4.3. Characterizing residential neighborhoods

In this study, we use census tracts to represent residential neighborhoods to demonstrate the model application. We made this choice mainly because the census tract is the most refined spatial scale shared among the collected datasets. Census tracts are created to include population groups of relatively homogeneous size (about 4,000 people) and characteristics with regularly measured social, demographic, and economic statistics (U.S. Census Bureau, 2017). Furthermore, the delineation of census tracts consults the built-environment and geographic features, e.g., rivers and major roads in local communities. These factors render census-based divisions well-suited as the basic analysis units for urban computing and modeling studies (e.g., Simini et al., 2012; Simini et al., 2021).

There are a total of 406 census tracts in the study region. For each census tract, we collected data characterizing its social-, economic-, and demographic profiles in addition to land use, physical accessibility, and hazard exposure as **Table 3** shows.

**Table 3**. Variables used to characterize residential neighborhoods

| Variable | Definition |
|---|---|
| Census characteristics | A 7-d vector depicting the social-, economic-, and demographic characteristics of the residential neighborhood:<br>• Age;<br>• Auto ownership;<br>• Below poverty ratio;<br>• Education;<br>• Employment status;<br>• Population size;<br>• Race (white vs. non-white). |
| Land use | An 8-d vector suggesting the proportion of areas of different land uses in the residential neighborhoods:<br>• Commercial;<br>• Mixed;<br>• Office;<br>• Parking;<br>• Recreation;<br>• Single-family residential;<br>• Multi-family residential;<br>• Vacant. |
| Accessibility | A 3-d vector suggesting the auto- and pedestrian-oriented accessibility including:<br>• Intersection density;<br>• Roadway density;<br>• Walkability. |
| Hazard exposure | A 3-d vector suggesting the proportion of areas of different flood zone types in the residential neighborhood including:<br>• X (low-risk 500-year floodplain);<br>• A (high-risk 100-year floodplain);<br>• V (coastal 100-year floodplain). |

### 4.4. Composing multi-modal transportation network

For each residential neighborhood and business cluster pair, we computed the travel time for households with and without vehicles referring to the multi-modal transportation networks in the study region. Specifically, we downloaded the transportation networks for driving and walking from the open street map (OSM) with the OSMnx package in Python (Boeing, 2017). We also downloaded the bus routes from the open GIS portal of the Miami metropolitan and processed them into the same format with the NetworkX package in Python (Hagberg et al., 2008).

We combined the transportation networks for different travel modes into a single network with NetworkX's *compose* function (**Figure 5**). The nodes are dead-end access points or street intersections while edges are roadway links characterized by allowed travel modes, length, and travel speed limits. To compute the travel time between two random nodes, we assigned speeds for vehicles and transits to be 50 km/h for roadways lacking speed limit information and 5 km/h for walking. We assumed that households with vehicles will drive to reach their destinations while households without vehicles will use the "transit + walking" mode. We then calculated the travel time for each roadway link under conditions with and without vehicles. We used the *shortest path* function in the NetworkX package to identify the shortest travel time connecting nodes that are closest to the centroids of residential neighborhoods and business clusters.

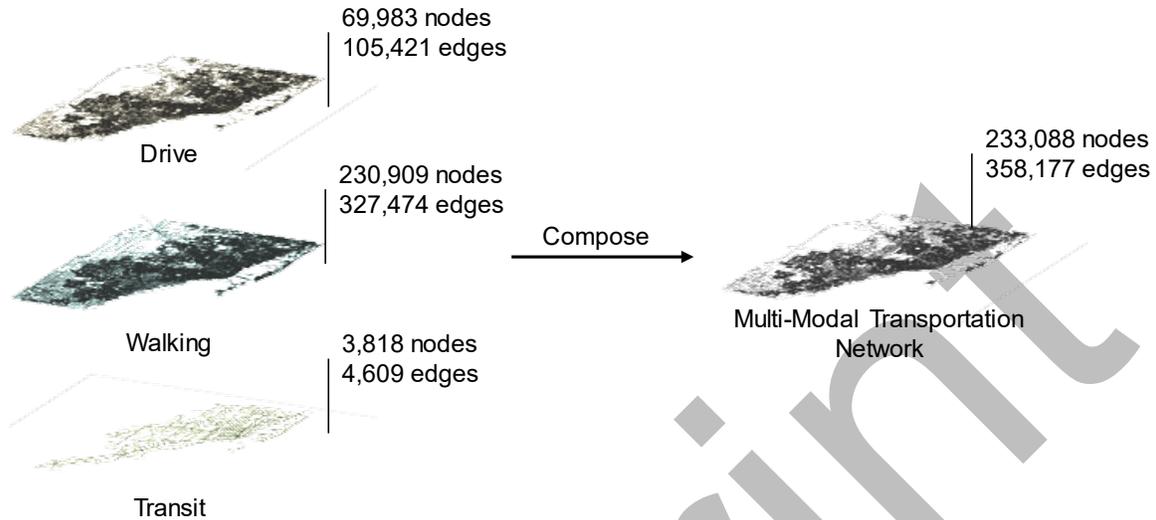

**Figure 5**. The composition of the multi-modal transportation network for the study region.

### 4.5. Characterizing dynamic environmental exposure

City residents' shopping behaviors are also influenced by environmental conditions. In the case of Miami, the region is especially susceptible to recurrent extreme events such as tropical storms and tidal flooding. We used a vector to characterize the different environmental exposure factors (**Table 4**). For weather conditions, we collected the daily weather records, including temperature, precipitation, and wind speed, using the Climate Data Online API (NOAA, n.d.a). Records of extreme events were sourced from the Storm Events Database (NOAA, n.d.b), which records various storm-related hazards or other significant weather phenomena that may affect community business clusters. We broadly categorized the varied hazards into coastal flood, flood, storm, and others (**Table 4**). The former three types of events account for 86.7% (98 out of 113) of records in the dataset. The category of "others" includes lighting, hail, wildfire, and heavy rains whose occurrences are too rare to be effectively captured by the model calibrated in this case study.

**Table 4**. Variables used to characterize the dynamic environment.

| Variable | Definition |
|---|---|
| Weather | A 5-d vector to represent the weather conditions of the metropolitan including:<br>• Average temperature;<br>• Average wind speed;<br>• Fastest 2-minute wind speed;<br>• Number of days included in the multiday precipitation;<br>• Precipitation intensity. |
| Hazard | A 4-d vector to represent days with the occurrence of the following hazard conditions:<br>• Coastal flood;<br>• Flood;<br>• Others (e.g., lighting, hail, and heavy rains);<br>• Storm (e.g., thunderstorm, tropical cyclone). |

| Covid-19 | We collected the period during which Miami-Dade County implemented the stay-at-home order and used a variable to mark the days corresponding to the periods with the stay-at-home order. |
|---|---|
| Holiday | We identified the national holidays during the study period and used a variable to mark the days corresponding to national holidays. |

## 5. Model Development

### 5.1 Model calibration

We pre-processed the data before the model development. We checked the distribution of the different variables that describe residential neighborhoods and business clusters and standardized them. For variables with long-tail distributions, such as the number of businesses in different business clusters, we applied a log transformation before standardizing the variables. This step ensured that each variable describing residential neighborhoods or business clusters followed a distribution close to the normal, binary, or uniform distribution.

We developed the SIM-GAT model with the data collected from the study region from January 1$^{st}$, 2019 to November 17, 2020. We randomly split the data into an 80% training set and a 20% validation set. The model was developed with the PyTorch framework on Python 3. We set the time length for the LSTM layer to be seven and the hidden dimensions for different layers to be 8. We used the Poisson loss function given that the outputs are "counts" of visitations following a Poisson distribution; 8 for the batch size; and 0.01 for the learning rate. This combination of hyperparameters was obtained with the grid search. The model was trained for 100 epochs and the epoch yielding the lowest loss in the validation set is stored for further analyses. Finally, the model achieved a Poisson loss of 1.4407 on the training set and 1.5555 on the validation set, respectively.

We trained the model to predict the daily visitation flows for residents from different residential neighborhoods to different business clusters based on dynamic inputs depicting the environmental and overall visitation conditions of the past seven days (**Table 4**). We show the spatial (averaging in temporal dimension) and temporal (aggregated in spatial dimension) predictions in **Figure 6** and **Figure 7**, respectively. The figures demonstrate the capability of our model to capture the spatial distribution of visitations among different clusters and temporal variances across different days.

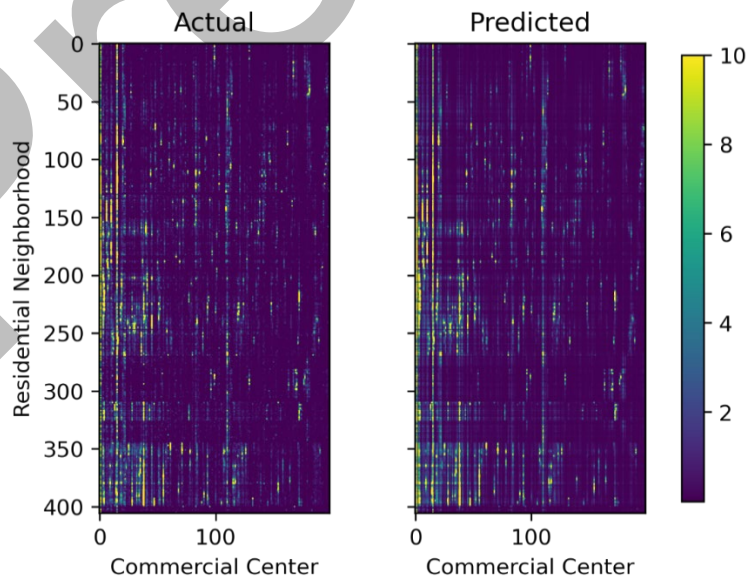

**Figure 6.** The spatial distribution of averaged actual and predicted visitation frequencies.

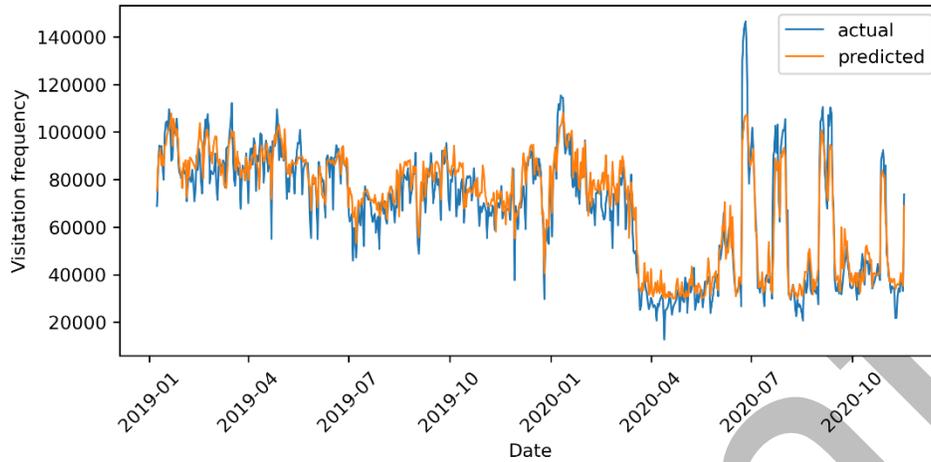

**Figure 7.** The temporal distribution of aggregated actual and predicted visitation frequencies.

### 5.2 Comparison with other graph-based deep learning model

We compared the performance of the proposed SIM-GAT model with two other graph-based deep learning models: GCN and GraphSAGE. Both models have gained increasing popularity in urban studies recently (Kipf & Welling, 2016; Hamilton et al., 2017). GCN is similar to a CNN layer that aggregates information from neighbors, but it is tailored for graph-structured data rather than grided data (Kipf & Welling, 2016). Unlike GAT, the adjacency matrix used in GCN is often pre-computed, typically reflecting spatial or semantic adjacencies of the graph nodes (Hao et al., 2023). GraphSAGE, on the other hand, learns node representations by aggregating information from neighboring nodes (Hamilton et al., 2017). This model can better capture the structural and relational information of the graph and is more suited for handling large graphs, such as social networks. Both GCN and GraphSAGE assume that nodes in the graph are of the same type, i.e., described with the same set of variables. Therefore, we used the graph with only business cluster nodes for the development of these two models.

In this study, we used the standard GCN and GraphSAGE architectures from the *PyTorch Geometric* library (Fey & Lessen, 2019). We kept other model layers and hyperparameters, including hidden dimensions, learning rate, number of training epochs, and the type of activation layers, the same as the calibrated SIM-GAT model. **Table 5** records the number of parameters and training/validation losses for the three models and **Figure 8** shows the learning curves for the three models. The proposed SIM-GAT model outperformed the GCN model and exhibited equivalent performance to the GraphSAGE model, however, with significantly fewer parameters (**Table 5**). This can be attributed to SIM-GAT's focus on learning localized interactions between business clusters and residential neighborhoods. The inferences of visitations are drawn from pair-wise interactions, while the other two models predict visitation flows for a business cluster with its own conditions and the conditions of nearby business clusters. The improved performance of SIM-GAT can also be partly attributed to the additional data used to describe residential neighborhoods – an aspect that the other two models cannot achieve.

**Table 5**. Performance metrics of SIM and SIM-GAT

| Model | # parameters | training loss | validation loss |
|---|---|---|---|
| SIM-GAT | 1,751 | 1.4407 | 1.5555 |
| GCN | 4,718 | 1.6502 | 1.7550 |
| GraphSAGE | 4,846 | 1.4392 | 1.5615 |

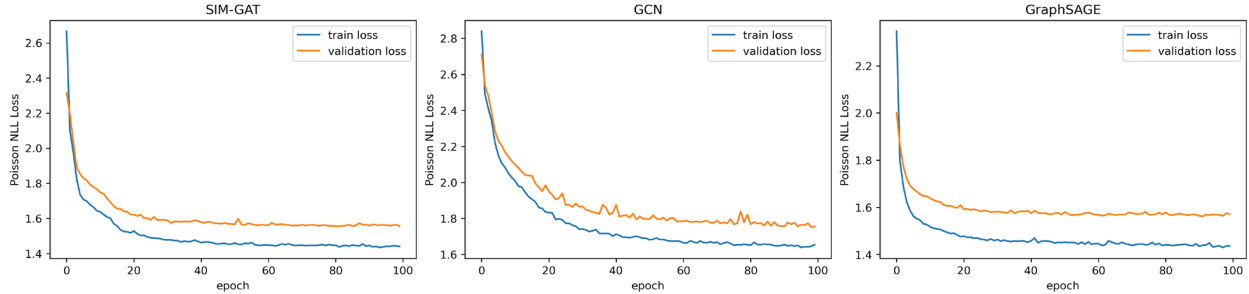

**Figure 8.** Learning curves for the three graph-based deep learning models.

## 6. Interpreting Context-Aware Attractiveness of Community Business Clusters with XAI

We used XAI to interpret the calibrated model and show how variables listed in **Table 2** influence the attractiveness of business clusters. In this study, we used the DeepLIFT, proposed by Shrikumar et al. (2017), in the *Captum* package (Kokhlikyan et al., 2020) for model interpretation. Compared to other XAI methods, DeepLIFT is particularly suited for handling non-linear activation functions (e.g., LeakyReLU), which were extensively used in the proposed SIM-GAT.

### 6.1 Variable contribution to business cluster attractiveness across scenarios

As the variable influence can vary across different business clusters and over time, we plotted the distributions of variable influences on business cluster attractiveness across scenarios as box plots in **Figure 9**. Specifically, we conducted XAI analysis for the 406 residential neighborhoods on a random normal day (i.e., January 8, 2019), the day Hurricane Dorian had the most severe impact on the study region (i.e., September 3, 2019), and one week after the Stay-at-Home order became effective in the region (i.e., March 25, 2020) to compare variable contributions to the attractiveness of business clusters across different scenarios. A positive value in the figures suggests that the presence or higher value of that variable contributes to more visitations to business clusters, i.e., yielding greater attractiveness, and vice versa. A variable with a box located tightly near zero suggests that the variable has minimum influences on business cluster attractiveness.

The XAI analysis has identified several variables that influence the attractiveness of business clusters. Specifically, large business clusters with more business establishments (*businessNum*) or larger areas (*totalArea*) are found to be more attractive (**Figure 9**). These findings are consistent with previous empirical studies (e.g., Teller & Reutterer, 2008, Dolega et al., 2016), which showed the positive effects of business cluster size on business cluster attractiveness. In terms of morphological structures, the XAI analysis results showed that business clusters located in downtown or commercial plazas are more attractive than commercial strips (**Figure 9**). Previous studies have also identified similar results that customers and planners prefer commercial nucleation (e.g., local or regional commercial districts) over commercial strips (e.g., Howe & Rabiega, 1992). Especially, commercial strips in the U.S. are mostly oriented towards automobiles, e.g., with rows of drive-thru fast restaurants, which is different from the pedestrian-oriented "high streets" in Europe. These commercial strips are less supportive for multi-purpose shopping or social functions, making them less attractive to customers. For flood vulnerability, the analysis showed that business clusters located in coastal areas, denoted with higher ratios of *fldZone_V*, and less flood-prone communities are more attractive compared to those located in in-land and flood-prone areas (i.e., denoted with higher ratio of *fldZone_A*) (**Figure 9**). For built-environment factors, it was shown that business clusters located near single- or multi-family residential areas (i.e., *lu_single-family* and *lu_multi-family*) are more attractive to local customers. Business clusters located in mixed-use areas or in areas with more commercial developments are comparatively less attractive. However, contrary to previous studies, the analysis showed that the contribution of public transportation accessibility (i.e., *busStopNum*) and business diversity (i.e., *businessDiversity*) to business cluster attractiveness vary drastically across different business clusters, as suggested by the wider box with a mean value close to zero, which may require further investigations.

**Figure 9** also presents the XAI analysis results for the two adverse conditions: Hurricane Dorian and the Stay-at-Home order during COVID-19. The results showed that the signs for variable contributions to business cluster attractiveness generally remain consistent with those obtained for normal conditions despite associated with a reduced magnitude (**Figure 9**). Especially, the XAI analysis for the Stay-at-Home order scenario showed that most variables depicting business clusters became not influential.

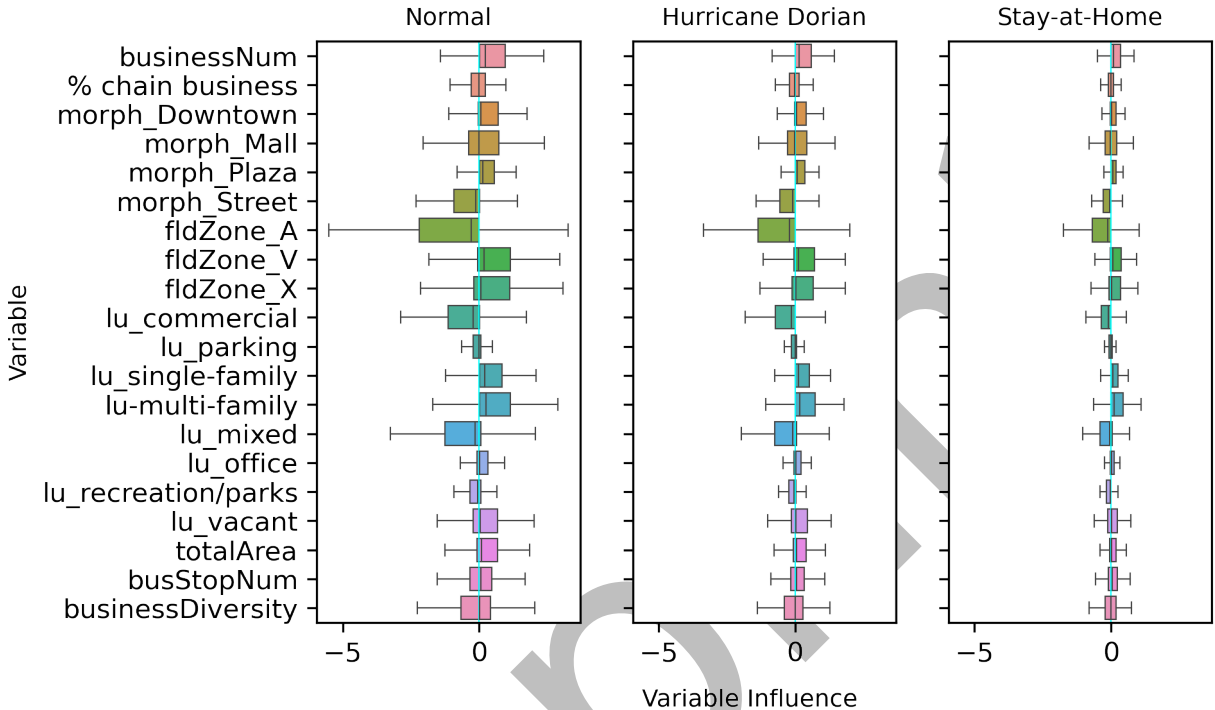

**Figure 9**. Variable contribution to the attractiveness of business clusters across scenarios.

### 6.2 Variable contribution to business cluster attractiveness across neighborhoods

We plotted the XAI analysis results for the influences of business cluster variables among different neighborhoods. We focused on a single normal day (i.e., January 8, 2019) and compared neighborhoods ranked top and bottom ten regarding per-capita income, auto-ownership rate, and the ratio of elderly residents (aged more than 65), respectively (**Figure 10**).

**Figure 10** shows that variables may exert different influences on business center attractiveness to different neighborhoods. For example, it was shown that residents in wealthy neighborhoods are more attracted to business clusters in downtown, taking the form of shopping malls, and located in coastal (i.e., measured with *fldZone_V*) and residential neighborhoods (i.e., measured with *lu_single-family* and *lu_multi-family*) (**Figure 10**). Similar trends are also identified for neighborhoods with higher auto-ownership rates, except that residents in neighborhoods with lower auto-ownership rates are even more attracted by business clusters located in multi-family residential areas (**Figure 10**). A possible explanation is that the multi-family residentials in Miami include both luxury condos and affordable apartments. For comparison, **Figure 10** shows that the variable influence on business cluster attractiveness does not differ much between neighborhoods with more or fewer elderly residents.

An interesting finding is that business clusters with more bus stops, measured with *busStopNum* are more attractive to low-income neighborhoods with lower auto-ownership rates. However, this variable has little influence on wealthy neighborhoods with high *per-capita income* or with a higher ratio of auto-ownership rate. This finding is plausible as low-income households without automobiles are more dependent on public transportation. **Figure 11** further shows that the increasing number of bus stops only positively influences business cluster attractiveness to neighborhoods with auto-ownership rates below 85%, but the influence is less clear for neighborhoods with higher auto-ownership rates.

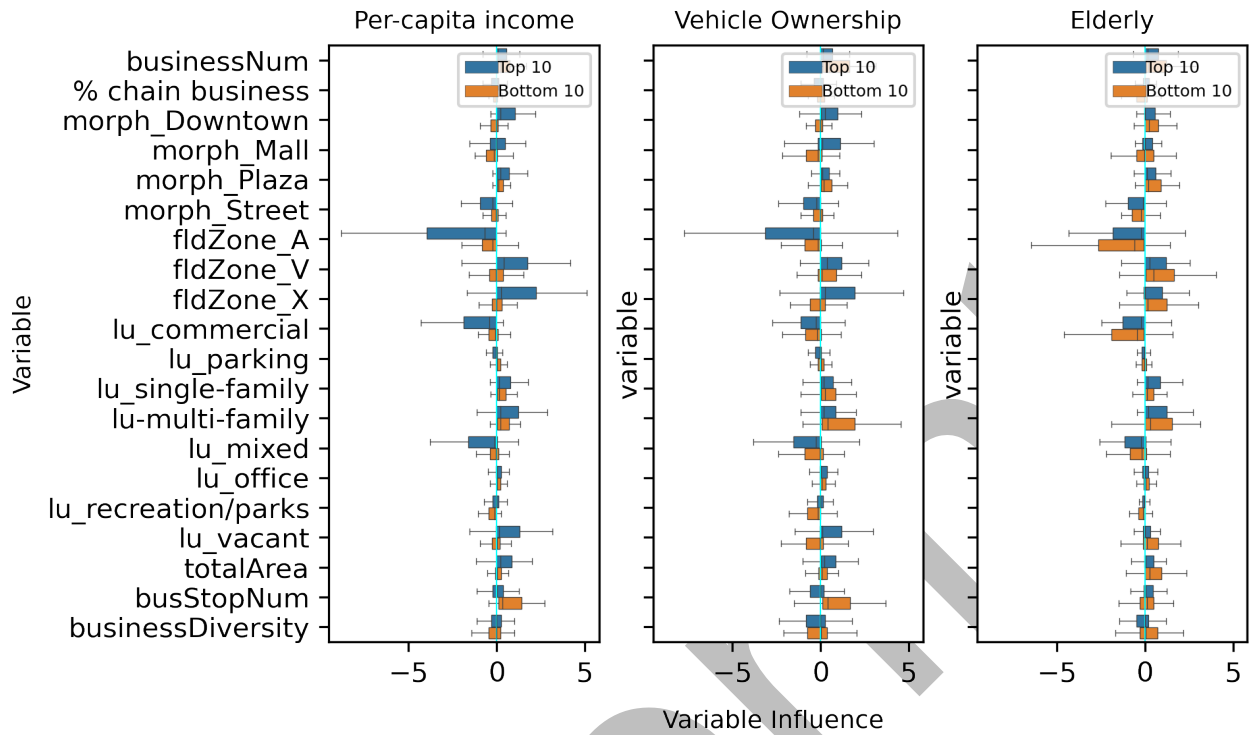

**Figure 10**. Variable contribution to the attractiveness of business clusters across residential neighborhoods.

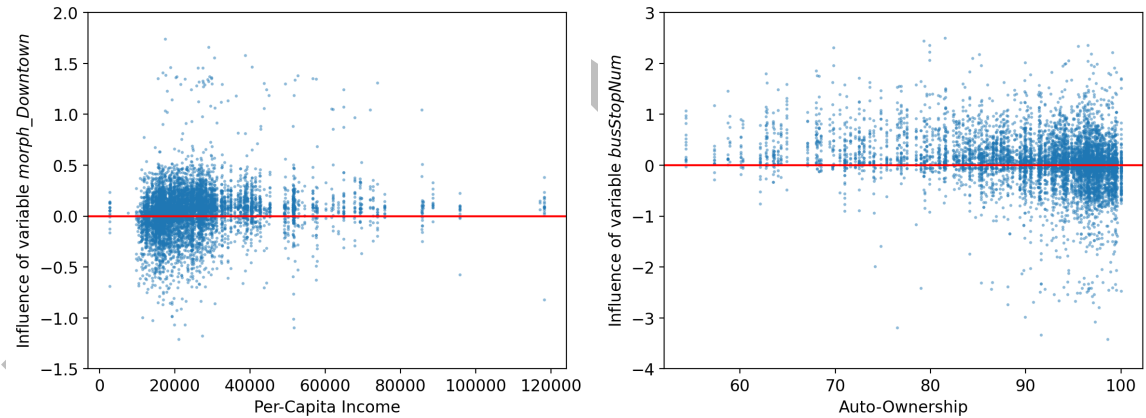

**Figure 11.** The influence of variable *morph_Downtown* distributed across neighborhoods with different per-capita income levels (left) and variable *busStopNum* distributed across neighborhoods with different auto-ownership rates (right).

### 6.3 Implication on resilience planning of community business clusters

Based on the XAI analysis, the following data-driven strategies are recommended to improve the resilience of community business clusters in Miami-Fort Lauderdale metropolitan. First, future commercial planning practices should prioritize developing nucleation-based commercial centers, such as commercial plazas, than ribbon-shaped commercial centers as the former morphology type is more attractive to local customers even during adverse events (**Figure 9**). Second, cities may discourage new businesses situated in vulnerable floodplains, i.e., associated with 1% annual probability of flooding, as business clusters located in those areas are associated with lower attractiveness to local customers. Third, cities may consider adding small or medium commercial clusters in neighborhoods dominated by single- or multi-residential developments, especially if those neighborhoods are also characterized by low auto-ownership rates (**Figure 9**). Forth, the planning of public transportation infrastructure, such as bus stops, supporting commercial clusters should be strategic and take account of the characteristics of surrounding neighborhoods. Throughout this process, computing methods like XAI can be leveraged to identify place-specific thresholds (e.g., 85% for auto-ownership rate in the study region) and supplement local planning practices (**Figure 11**).

## 7. Discussion

In this study, we proposed SIM-GAT, a model coupling the conventional SIM with a graph-based deep learning model to capture the dynamics of community business clusters. We have developed and calibrated the model with data collected from the Miami metropolitan area in Florida and analyzed the varied variable contributions across scenarios and among outlet groups. Based on the modeling and analysis results, the following contributions and implications are discussed.

First, we propose a deep learning-based method to complement the conventional SIM for modeling the dynamics of community business clusters. The method innovatively represents the interconnected business clusters within an urban region as a graph consisting of nodes and edges, which include components of business clusters, trade areas, and multi-modal transportation infrastructure. This graph representation provides a necessary abstraction to the complex system of community business clusters, allowing the method to account for intricate spatial correlations among different business clusters and outlet trade areas that are overlooked by conventional approaches (LeSage & Pace, 2008; Suárez-Vega et al., 2015). The use of deep learning also empowers the model in several ways. Particularly, previous studies have found it difficult to measure the aggregated attractiveness of individual retail centers (Reutterer & Teller, 2009; Teller & Reutterer, 2008). We address this challenge by treating the attractiveness measures for business clusters as feature vectors, depicting the characteristics of community business clusters, and using attention and non-linear activation functions in the SIM-GAT model to account for the interactions between business clusters and residential neighborhoods. This design can better account for the non-linearity and context-aware nature of the complex system of community business clusters compared to the addition, regression, or multiplication approaches used in previous studies (Dolega et al., 2016; Nakanishi & Cooper, 1974; Sevtsuk, 2014).

Second, the XAI analysis has shown the varying variable contribution to the attractiveness of community business clusters across scenarios and among different residential neighborhoods. Residents in Miami-Fort Lauderdale generally favor two types of business clusters, i.e., large and diverse business centers in busy downtown and local plazas near residential neighborhoods. While most of our findings are consistent with previous studies, our use of a deep learning model calibrated with local data also supplements data-driven and place-specific evidence. For example, we have shown that adding bus stops to business clusters serving residential neighborhoods with auto-ownership rates lower than 85% is more contributive in the study region. Such nuanced findings, i.e., place-specific thresholds, can not be obtained with previous methods, but can inform strategic planning practices improving the attractiveness and resilience of community business clusters across scenarios.

Our research has several limitations that open opportunities for future studies. Like many other urban computing and modeling studies, the presented case study for illustrating model application is limited by data availability and granularity. One limitation is that the SafeGraph platform does not collect the detailed origins of each trip. We therefore assumed that the origins of the trips are visitors' home census tracts, which may lead to biased analysis results. According to the 2017 National Household Travel Survey (NHTS) data (FHA, 2017), home-based shopping trips consist of 52.4% of total sampled shopping trips. For non-home-based shopping trips, the prior activities are mostly for other shopping purposes (53.9%) or working (11.0%). The assumption made in the case study may underestimate the demands for neighborhoods with more office and commercial developments. Second, we used several factors to indicate the presence of hazard conditions in the study region. However, there is a lack of fine-grained data showing the spatial distribution of the hazard conditions throughout the event, such as inundation maps. Future studies may consider integrating the outputs of numeric models or simulation software to address this limitation. Third, we used visitation counts for the response variable, given that customers "vote with their foot" for the attractiveness of different business clusters. However, visitations alone may not tell a whole story, especially considering the increasing uptaking of delivery orders. To address this, we have explored SafeGraph's spending data, which includes transaction records from e-commerce platforms such as DoorDash and Postmates. Unfortunately, the e-commerce data collected by SafeGraph do not reveal spatio-temporal patterns like the visitation data. Future studies could incorporate transaction and e-commerce data in our proposed model, e.g., through multi-task learning, for a more comprehensive understanding of business cluster attractiveness. Moreover, the inclusion of the GAT layer provides theoretical support for the applicability and generalizability of SIM-GAT to other cities. However, we acknowledge that business dynamics can vary drastically across cities that are different in terms of size, geography, and culture. We are planning future research to explore the feasibility of creating a "general" SIM-GAT model working for different cities and identifying the needed resources for such knowledge transferring.

## 8. Conclusion

Communities around the world are facing increasing challenges posed by climate change, pandemics, economic crises, and other sources of adversities. We introduce a deep learning-based approach, i.e., SIM-GAT, to assist in the analysis and planning of inter-connected community business clusters. The model can capture the spatio-temporal dynamics of community business clusters across scenarios. We have demonstrated the effectiveness of the model with data collected from the Miami Metropolitan area in Florida. The interpretation of the calibrated model using XAI has confirmed the critical role of planning factors in shaping the dynamics of community business clusters. Such a data-driven model and findings of XAI can be leveraged by urban planners or investors to create more resilient and tailored commercial plans that can withstand various stressors and accommodate community residents. The model can also be used in scenario planning practices by analyzing the performances of candidate commercial development plans across scenarios and help cities better counter adverse events. Furthermore, the study demonstrates the potential of AI in transforming conventional urban models by providing more accurate and detailed insights into the complex human-environment interactions that shape urban systems and dynamics.

**Acknowledgment:**

This material is based on work supported by the National Science Foundation under Grant No. 2316450 and Graduate School Fellowship Award at the University of Florida. Any opinions, findings, and conclusions or recommendations expressed in this material are those of the authors and do not necessarily reflect the views of the National Science Foundation and the University of Florida.